\documentclass[aip,rsi,amsmath,amssymb,reprint]{revtex4-1}

\usepackage{graphicx}% Include figure files
\usepackage{dcolumn}% Align table columns on decimal point
\usepackage{bm}% bold math
\begin{document}

\title[Kinetic energy and momentum distribution of isotopic liquid helium mixtures]{Kinetic energy and momentum distribution of isotopic liquid helium mixtures}

\author{Massimo Boninsegni}
\email{m.boninsegni@ualberta.ca}
\affiliation{ 
Department of Physics, University of Alberta, Edmonton, Alberta, Canada T6G 2G7
}%

\date{\today}

\begin{abstract}
The momentum distribution and atomic kinetic energy of the two isotopes of helium in a liquid mixture at temperature $T$=2 K are computed by quantum Monte Carlo simulations. Quantum statistics is fully included for $^4$He, whereas $^3$He atoms are treated as distinguishable. Comparison of theoretical estimates with a collection of the most recent experimental measurements  shows reasonable agreement for the energetics of $^4$He and pure $^3$He. On the other hand, a significant  discrepancy (already observed in previous works) is reported between computed and measured values of the $^3$He kinetic energy in the mixture, especially in the limit of low $^3$He concentration. 
We assess quantitatively the importance of Fermi statistics and find it to be negligible for a $^3$He concentration $\lesssim 20\%$.
Our results for the momentum distribution lend support to what already hypothesized by other authors, namely that the discrepancy is likely due to underestimation of the $^3$He kinetic energy contribution associated with the tail of the experimentally measured momentum distribution. 
\end{abstract}
\maketitle

\section{\label{intro}Introduction}

Liquid mixtures of the two isotopes of helium have long been regarded as an interesting playground for quantum many-body physics.\cite{ahlers75} For example, in the limit of low $^3$He concentration ($x$), in which it remains homogeneous as the temperature $T\to 0$, such a mixture is perhaps the cleanest and most easily controlled experimental realization of a Bose superfluid ($^4$He) in the presence of mobile impurities. In that limit, $^3$He behaves very nearly as an ideal, essentially non-interacting Fermi gas whose degeneracy can be tuned by varying $x$. As $x$ is increased, both the interaction of $^3$He quasiparticles and the effect of $^3$He Fermi statistics become  more and more significant, as quantitatively expressed by higher values of the Fermi momentum and temperature. Indeed, Fermi statistics decisively contributes to shaping the experimental phase diagram of the mixture at temperatures below $\sim$ 1 K.  And because the interaction between two helium atoms is very nearly independent of spin and nuclear mass, an isotopic helium mixture is also an ideal system in which nuclear quantum effects can be studied.\cite{bbp}
\\ \indent
State-of-the art theoretical calculations based on realistic interatomic potentials have provided considerable qualitative and quantitative insight into the physics of the mixtures. For example, Path Integral Monte Carlo (PIMC) simulations\cite{bc,bm} have yielded definite predictions for the effective mass and chemical potential of a single $^3$He atom dissolved in superfluid $^4$He, and quantitatively reproduced  the experimentally observed, monotonic decrease of the $^4$He superfluid response with increased $x$, in the miscibility region. 
\\ \indent
The same agreement between theory and experiment has been lacking, however, for the single-particle atomic kinetic energy, which can be obtained as the second moment of the momentum distribution $f_\alpha({\bf k})$, $\alpha=3,4$, in turn measurable by means of neutron scattering experiments.\cite{wangandsokol,azuah95,senesi03,andreani06} While there is reasonable quantitative agreement between the experimental and theoretical estimates of the $^4$He kinetic energy per atom ($K_4$), reported values of the corresponding $^3$He quantity ($K_3$) have been consistently below the theoretical ones, by amounts worth as much as several K (peaking at around 50\% of the experimental value in the $x\to 0$ limit), well outside the quoted statistical uncertainties. Additionally, and perhaps even more importantly, while all the most reliable theoretical results show a clear monotonic decrease of $K_3$ on $x$ at low $T$, experimental data show virtually no dependence of $K_3$ on $x$, despite the substantial ($\sim$ 30\%) difference in equilibrium density between the $x=0$ (pure $^4$He) and $x=1$ (pure $^3$He) limits; this surprising observation was made in different independent measurements.\cite{azuah95,senesi03,andreani06} 
\\ \indent 
A discrepancy of this magnitude for a quantity like the kinetic energy, in a relatively simple system like the one considered here, could possibly point to some significant gap in the present understanding of the physics of the mixture, specifically the local environment experienced by a single $^3$He atom dissolved in superfluid $^4$He.
It was suggested in Ref. \onlinecite{andreani06} that the disagreement may point to ``effects of Fermi statistics'' as the (unexplained) cause of the departure from the expected density dependence of the single-particle mean kinetic energy. While it is certainly true that the Fermi component of the mixture is that for which the disagreement between theory and experiment is quantitatively most important, one is hard pressed thinking of a physical mechanism underlain by Fermi statistics whose overall result would be that of {\em lowering} the kinetic energy. This seems especially the case in the low $x$ limit, and at a temperature as high as $T$=2 K,  where effects of Fermi statistics should be relatively small. For example, assuming a $^3$He effective mass of the order of 2.3 times the bare mass\cite{bc} one can estimate the degeneracy temperature of the $^3$He fluid in the mixture for $x=0.1$ at its equilibrium density to be $\lesssim 0.5$ K.
\\ \indent
An alternate explanation\cite{bc,diallo06} is that the root of the discrepancy may lie in the (possibly large) contribution to the kinetic energy per particle from the tail of the momentum distribution, the estimate being quite sensitive to the model function utilized to fit the experimental data, particularly for the Fermi component. Specifically, it was contended\cite{diallo06} that the experimental underestimation of the $K_3$ may stem from the use of a free Fermi gas type model for the $n_3({\bf k})$, inadequate to describe the significant high-momentum tail of the observed distribution. Indeed, is was proposed therein that a more reliable comparison may be between the calculated and observed $n_3({\bf k})$ rather than their second moment (namely $K_3$) which is not well determined experimentally. 
\\ \indent
This is the kind of quantitative, well-defined questions that computer (QMC) simulations can usually address effectively. Unfortunately, numerically exact results for the $n_3$ at finite $x$ are difficult to obtain, due to the well known fermion ``sign'' problem, plaguing any quantum Monte Carlo technique, including PIMC. In Ref. \onlinecite{bm}, use was made of the so-called restricted path integral (RPIMC) technique,\cite{ceperley92} which removes the sign instability at the cost of making an uncontrolled approximation, namely restricting paths to regions in which a trial many-fermion density matrix (in this specific case that of a system of free fermions) is positive. The use of this approximation can be justified in the $x\to 0$ limit, in which $^3$He should behave as an ideal Fermi gas. On the other hand, in the opposite (pure $^3$He) limit the results afforded by this approach are only semi-quantitative.\cite{ceperley92} In any case, no RPIMC results have been reported to date of the momentum distribution of any Fermi system. 
\\ \indent 
At least at $T$=2 K, however, it is conceivable that one may obtain reliable results by neglecting $^3$He Fermi statistics altogether, i.e., by regarding $^3$He atoms as {\em distinguishable}. This approximation removes the sign problem and allows one to compute by QMC the momentum distribution for both components, affording a direct comparison of theoretical and experimental results. This is the computational strategy adopted in this work.
\\ \indent 
This paper reports results of QMC simulations of the mixture in the $0\le x \le 1$ range, at temperature $T$=2 K, although a few simulations at $T$=1 K were carried out as well for comparison. The main physical quantities of interest are the momentum distributions $f_\alpha({\bf k})$ and the atomic kinetic energies $K_\alpha(x)$. Quantum statistics is fully included for the Bose component, namely $^4$He, whereas as stated above quantum exchanges are excluded for the $^3$He fluid (i.e., $^3$He atoms are assumed to obey Boltzmann statistics). Quantitative arguments are furnished to the effect that this is indeed an excellent approximation at $T$=2 K and for $x\lesssim 0.2$.
\\ \indent 
The results confirm the disagreement between theoretical and experimental estimates for $K_3$ at low $x$, while  for $K_4$ the agreement with experiment in the whole $x$ range, while not impressive, seems satisfactory. It is worth mentioning that the  calculation carried out here yields a kinetic energy values in reasonable agreement with experiment at $x=1$, i.e., for pure $^3$He, where effects of Fermi statistics should be most important. Actually, the total energy value computed at this temperature for pure $^3$He at its equilibrium density seems to be in closer agreement with experiment than the RPIMC one from Ref. \onlinecite{ceperley92}. Whether that is the result of a fortuitous compensation of error, or whether maybe it points to exchanges being less important in fluid $^3$He at $T$=2 K than previously thought is unclear, but certainly worthy of further investigation.
\\ \indent
The one-body density matrix for the $^3$He component in the $x\to 0$ limit displays an exponential tail at long  distances, which is consistent with the picture of a $^3$He atom dissolved in superfluid $^4$He as penetrating a potential barrier as it moves past the surrounding $^4$He atoms.  The $^3$He momentum distribution deviates significantly from a Gaussian, displaying both a low momenta enhancement, as well as a slowly decaying tail at high momenta, altogether supporting the contention of Ref. \onlinecite{bc,diallo06}, and suggesting that the disagreement between theoretical and experimental estimates may be removed by the use of an appropriate model for $n_3$, featuring a high momentum tail, to fit the experimental data.
\\ \indent 
The remainder of this paper is organized as follows: in Sec. \ref{mm} the model of the system is introduced, and the computational methodology briefly reviewed; the results are presented in detail in Sec. \ref{res}; conclusions are outlined in Sec. \ref{conc}.

\section{model and methodology}\label{mm}
The mixture is described as an ensemble of $N$ pointlike particles, of which $Nx$ are $^3$He, which are regarded as distinguishable, the rest $N(1-x)$ $^4$He atoms, obeying Bose statistics. The system is enclosed in a cubic cell with periodic boundary conditions in the three directions. 
\\ \indent 
The quantum-mechanical many-body Hamiltonian of the system reads as follows:
\begin{eqnarray}\label{u}
\hat H = -\sum_{i\alpha}\lambda_{\alpha}\nabla^2_{i\alpha}+\sum_{i<j}v(r_{ij})
\end{eqnarray}
where the first sum runs over all particles of either species,with $\lambda_{3}\ (\lambda_{4})=8.0417 \ (6.0596)$ K\AA$^{2}$, whereas the second sum runs over all pairs of particles, $r_{ij}\equiv |{\bf r}_i-{\bf r}_j|$ and $v(r)$ is the accepted Aziz pair potential,\cite{aziz79} which describes the interaction between two helium atoms of either species. Such a potential ha been shown to afford a rather accurate description of the energetic and superfluid properties of $^4$He. In principle a more accurate model
would go beyond the simple pair decomposition, including,
for instance, interactions among triplets; however, published
numerical work has given strong indications
that three-body corrections, while significantly affecting the
estimation of the pressure, have a relatively small effect on
the structure and dynamics of the system, of interest here.\cite{mpfb}
\\ \indent 
The low temperature phase diagram of the system described by Eq.  (\ref{u}) as a function of $x$ has been studied in this work by means of  first principles numerical simulations, based on the continuous-space Worm Algorithm. \cite{worm,worm2}  Since this technique is by now fairly well-established, and extensively described in the literature, we shall not review it here. A canonical variant of the algorithm was utilized, in which the total number of particles $N$ is held fixed. \cite{mezz1,mezz2} As mentioned above, $^3$He atoms are regarded as distinguishable, whereas $^4$He atoms obey Bose statistics; however, for the purpose of gaining further insight a few simulations were performed in which $^3$He atoms were treated as {\em Bosons}, with the inclusion of quantum-mechanical exchanges. This point will be discussed in depth in Sec. \ref{res}. 
\\ \indent 
Details of the simulation are  standard; for instance,  the short-time approximation to the imaginary-time propagator used here is accurate to fourth order in the time step $\tau$ (see, for instance, Ref. \onlinecite{jltp}).  All of the results presented here are extrapolated to the $\tau\to 0$ limit; in general, it was found that numerical estimates for structural and superfluid properties of interest here, obtained with a value of the time step $\tau = (1/640)$ K$^{-1}$ are indistinguishable from the extrapolated ones, within the statistical uncertainties of the calculation.
We carried out simulations of mixtures comprising   $N=256$ particles altogether. 
\\ \indent 
The physical quantity of interest, besides the usual energetic and structural ones, as well as the $^4$He superfluid fraction (computed using the well-known winding number\cite{pollock87} estimator), is the one-body density matrix
\begin{eqnarray}\label{obdm}
n_\alpha({\bf r},{\bf r}^\prime)=
\langle\hat\psi^\dagger_\alpha({\bf r}^\prime)\ \hat\psi_\alpha({\bf r})\rangle
\end{eqnarray}
where $\langle\cdots\rangle$ stands for thermal expectation value, and $\hat\psi_\alpha,\hat\psi^\dagger_\alpha$ are field operators for the two components. For a translationally invariant  and isotropic system like a homogeneous fluid, it is $n_\alpha({\bf r},{\bf r}^\prime)\equiv n_\alpha(|{\bf r}-{\bf r}^\prime|)$. The one-body density matrix is easily accessible for both components, using the worm algorithm. The momentum distribution is obtained as a three-dimensional Fourier transform, namely
\begin{eqnarray}\label{nofk}
f_\alpha({\bf k})\equiv f_\alpha(k)=\frac{4\pi}{k}\ \int_0^\infty dr\ r\ {\rm sin}(kr)\ n_\alpha(r)
\end{eqnarray}
with the normalization
\begin{equation}\label{norm}
\frac{1}{(2\pi)^3}\ \int d^3k \ f_\alpha({\bf k}) = 1
\end{equation}
which fixes to unity the value of $n_\alpha(r)$ at the origin.
\section{results}\label{res}
\begin{table}
\caption{\label{table1}Theoretically computed (columns marked with (T)) kinetic energy per particle $K_3(x)$ of Boltzmann  $^3$He, and $K_4(x)$ of Bose $^4$He for mixtures of different concentration $x$, at a temperature $T$=2 K. Results are in K. Included are also experimental results (columns marked with (E)) from Refs. \onlinecite {wangandsokol,azuah95,senesi03,andreani06}, as reported in Ref. \onlinecite{andreani06} (Table I). Statistical uncertainties, in parentheses, are on the last digit(s).  The results at $x=0$ (100)\% refer to a single $^3$He ($^4$He) atom in bulk $^4$He ($^3$He). }
\begin{ruledtabular}
\begin{tabular}{cccccc}
$x (\%)$ &$\rho$ (\AA$^{-3}$) & $K_3$ (T) & $K_3$ (E) &$K_4$ (T) &$K_4$ (E) \\
0 &0.02187 &18.4 (2)&& 15.0(1) &16.0(5)\\
10 &0.02140 &17.96(8) &12.1(4) &14.87(6) &13.8(6)\\
20 &0.02090 &17.54(6) &10(2) &14.89(4) & \\ 
35 &0.01995 &16.56(8) &10.4(3) &14.20(5) &12.0(6) \\
65&0.01822 &14.52(4) &11.8(7) &12.62(4) & \\
100 &0.01550  &12.42(6) &12(1)&10.0(1) &
\end{tabular}
\end{ruledtabular}
\end{table}
Our results for the kinetic energy per helium atom $K_\alpha(x)$ at a temperature $T$=2 K, are shown in Table \ref{table1}. The values of the density of the liquid mixture at which calculations were carried out are taken from Table I of Ref. \onlinecite{andreani06}. It is worth mentioning again that at this temperature the mixture is {\em homogeneous}, i.e., no phase separation takes place at any $x$. The results presented in Table \ref{table1} are consistent with those of the previous calculations,\cite{bc,bm} taking into account slight differences in density, for $^3$He concentrations below $\sim$ 20\%. For higher $x$, the $^3$He kinetic energy is underestimated in this work, as effects of quantum statistics become important; for example, for $x=35$\% the value of $K_3$ reported here is as much as 1 K lower than that of Ref. \onlinecite{bm}, in which Fermi statistics is at least in part included through the nodal restriction. On the other hand, the $^4$He kinetic energy obtained here is consistent with that of Ref. \onlinecite{bm}.
\\ \indent 
The first immediate observation is the large discrepancy between theoretical and experimental estimates of $K_3$, in line with what reported in previous works. The deviation is largest in the limit $x\to0$, whereas in the opposite limit, i.e., $x\to 1$, the theoretical estimate for $K_3$ is actually consistent with experiment, obviously making allowance for the relatively  large uncertainty quoted in Ref. \onlinecite{andreani06}. For $^4$He the agreement is better but not entirely satisfactory either; specifically, in at least one case ($x$=35\%) the difference between experimental and theoretical estimate is well outside the quoted statistical uncertainties.
\\ \indent 
The limit of low $^3$He concentration $x$ is that in which the disagreement between theory and experiment, regarding the quantitative determination of $K_3$, is greatest in magnitude, and as mentioned above the suggestion was made that poorly understood effects of Fermi statistics may be responsible for it.\cite{andreani06} Because in this work exchanges of indistinguishable $^3$He atoms are neglected, it seems appropriate to offer a quantitative justification for this approximation, which is crucial in order to carry out the QMC simulation without incurring into the `sign" problem.
One may begin by noting that, in order for exchanges of identical particles to occur sufficiently frequently, the characteristic spatial extension $\Lambda_T\equiv (2\lambda/T)^{1/2}$ of a single-particle ``path'' should be of the order of the average distance $d$ between two such particles. For, the relative probability for an exchange including $n$ particles to take place, is roughly proportional\cite{rpf53} to $\gamma^n$, where $\gamma={\rm exp}[-d^2/\Lambda_T^2]$.
Consider for definiteness a $x=20\%$ mixture at a temperature $T$=1 K; assuming an unpolarized $^3$He component, the average distance of two $^3$He atoms with parallel spin projections is $\sim 7.8$ \AA; on the \textbf{}other hand, $\Lambda_T\sim 2.8$ \AA, if a $^3$He effective mass equal to twice the bare mass is assumed; thus, $\gamma\sim 5\times 10^{-4}$, which can be compared, for example, to the value $\sim 10^{-1}$ for liquid $^4$He at the superfluid transition temperature. Thus, one may expect exchanges of $^3$He atoms to be {\em strongly} suppressed in this system, at least down to this temperature. A similar analysis shows that this conclusion holds {\em a fortiori} for mixtures with lower $x$. 
\\ \indent 
Direct, quantitative support for the above conclusion is offered by simulations of a fictitious mixture in which {\em both} $^3$He and $^4$He atoms are assumed to be spin-zero Bosons, with exchanges allowed for both components.\footnote{It must be made clear that this exercise is only aimed at estimating the importance of permutation in the mixed Bose-Fermi system; it is well known that the physics of Bose mixtures is qualitatively different, though specific physical issues, e.g., phase separation, can be addressed in a Bose mixture as well (see, for instance, M. Boninsegni, Phys. Rev. Lett. {\bf 87}, 087201 (2001)).} At $T$=2 K and $x=20\%$ it is found that exchanges of Bose $^3$He atoms are exceedingly infrequent; specifically, over 99\% of all single-particle paths close onto themselves, and this percentage remains above 90\% as the temperature is lowered to 1 K. Moreover, the rare exchanges that occur mainly involve relatively few particles (of the order of five). As expected, exchanges occur even more infrequently at lower $^3$He concentration. \footnote{It should also be noted that this overestimates the frequency of occurrence of permutations in the fermion system, because in the case of spin-zero bosons exchanges of all particles are allowed, whereas in the case of an unpolarized spin-1/2 Fermi system like $^3$He, only half of the particles are potential exchange mates.} Because exchanges of identical particles are sampled exactly in the same way in Fermi or Bose systems (the difference being rather in how contributions to physical observables associated to exchange paths are added to averages), one can {\em a fortiori} conclude that effects of Fermi statistics in an isotopic helium mixture are  negligible for concentrations below $\lesssim 20\%$ at least down to temperature $T$=1 K; in other words, regarding $^3$He atoms as distinguishable particles is an excellent approximation. Thus, one may confidently expect that  in this region of the phase diagram, estimates of most structural and energetic properties of the mixtures computed in this way should be fairly accurate. However, special care must be exercised when it comes to the one-body density matrix and the momentum distribution, which, as the example of liquid parahydrogen near freezing  shows, can display important signatures of quantum statistics, virtually  undetectable in all other observables.\cite{mb09}
\\ \indent
At greater $^3$He concentrations effects of Fermi statistics are expected to become increasingly important; curiously, however
the disagreement between theory and experiment regarding $K_3$ is quantitatively smaller in this limit. For example, for pure liquid $^3$He at $T$=2 K the atomic kinetic energy at the experimental density quoted in Ref. \onlinecite{andreani06} is consistent with experiment. It is also worth mentioning that a calculation carried out in this work at density $\rho=0.016355$ \AA$^{-3}$ for pure liquid $^3$He at $T$=2 K yields an energy per particle equal to -1.51(4) K, actually in rather good agreement with experiment, at least comparable to (if not better than) that afforded by the original RPIMC calculation\cite{ceperley92} for normal $^3$He, yielding approximately -1.3 K. While this could be the result of a fortuitous compensation of error,\footnote{It is known that the ground state of a system of distinguishable particle is the same with that of the corresponding Bose system; thus, it is to be expected that energies computed by neglecting Fermi statistics will at low temperature fall {\em below} those of the Fermi system. Indeed, the same calculation at $T$=1 K yields an energy per $^3$He atom close to -2.4 K, about 0.4 K below the experimental value. For a more extensive discussion see, for instance, M. Boninsegni, L. Pollet, N. Prokof'ev and B. Svistunov, Phys. Rev. Lett. {\bf 109}, 025302 (2012). } it suggests that quantum exchanges may be perhaps quantitatively less important than expected for this system, at this temperature. One way to test this hypothesis may be that of incorporating the $^3$He effective mass enhancement in the nodal restriction of the  calculation of Ref. \onlinecite{ceperley92} (also based on the free Fermi gas approximation); this has the result of suppressing in part exchanges, because of  the shorter atomic thermal wavelength.

\begin{figure}[ht]
\centering
\includegraphics[width=\linewidth]{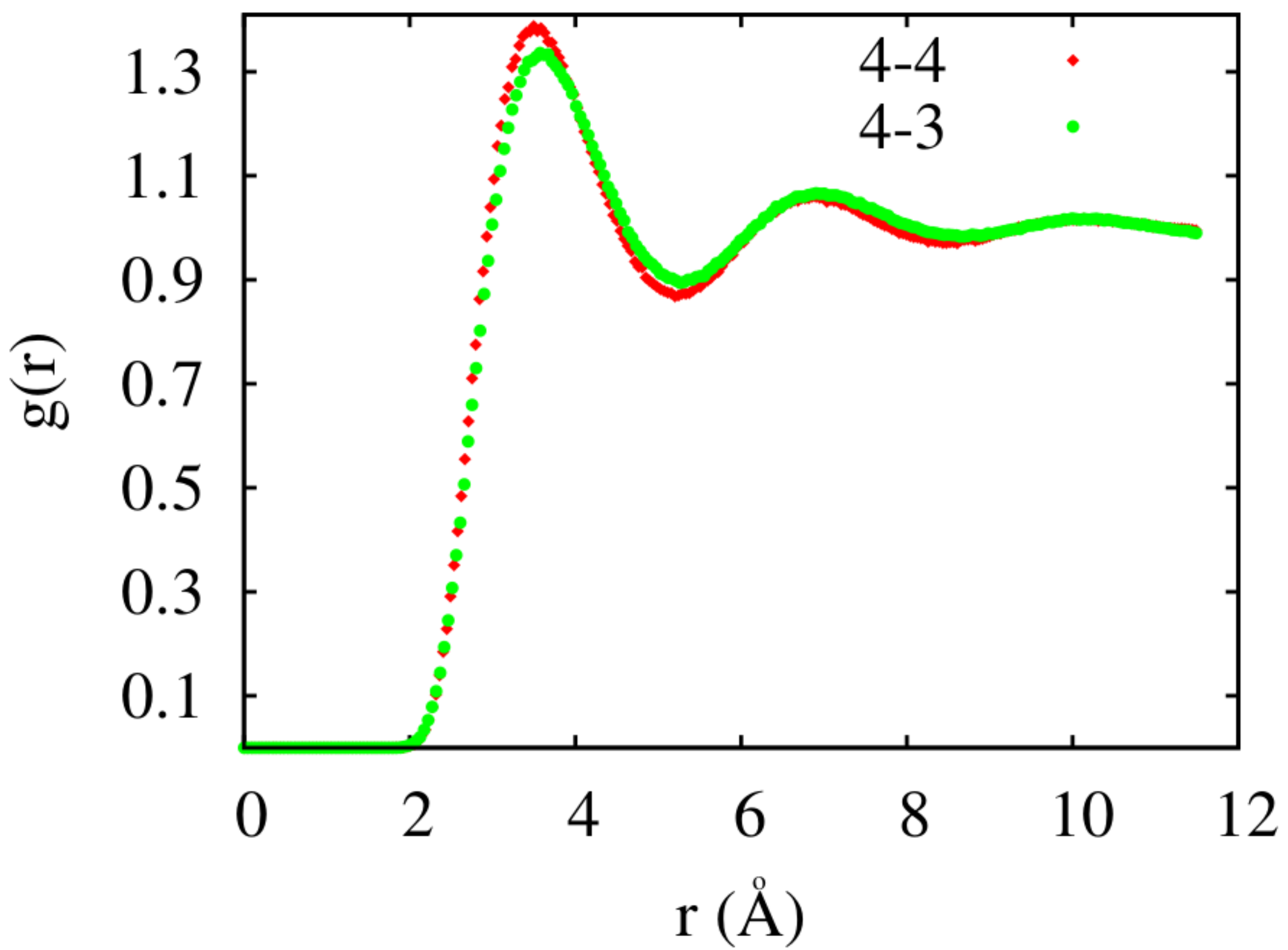}
\caption{{\rm Color online.} Pair correlation functions $g(r)$ for a liquid helium mixture with a $^3$He concentration $x=10\%$ at  temperature $T$=2 K, computed by QMC simulation. The density of the mixture is 0.0214 \AA$^{-3}$. Shown are the results for the $^4$He-$^4$He (triangles) and $^4$He-$^3$He correlation functions. Statistical errors are smaller than the sizes of the symbols.}
\label{f1}
\end{figure}

We now illustrate our results for a mixture with $x=10\%$, for which the disagreement between experimentally determined and theoretically computed $K_3$ is rather large (Table \ref{table1}).
Fig. \ref{f1} shows the pair correlation functions $g(r)$ computed at $T$=2 K, both that for two $^4$He atoms as well as that between a $^4$He and a $^3$He atom; although there are some detectable differences, it is clear that the local environment experienced by a $^3$He atom in the mixture is essentially the same as that experienced by a $^4$He atom. There is no evidence that $^3$He atoms push $^4$He atoms further away, in order to reduce their kinetic energy of confinement, as speculated, for instance, in Ref. \onlinecite{andreani06}. Thus, since the interatomic potential is the same for all pairs, and since as stated above effects of Fermi statistics are negligible, one can account for the $^3$He kinetic energy increase with respect to the pure $^3$He case (an increase of roughly roughly 6 K) simply  based on the higher equilibrium density of the mixture. 
\begin{figure}[ht]
\centering
\includegraphics[width=\linewidth]{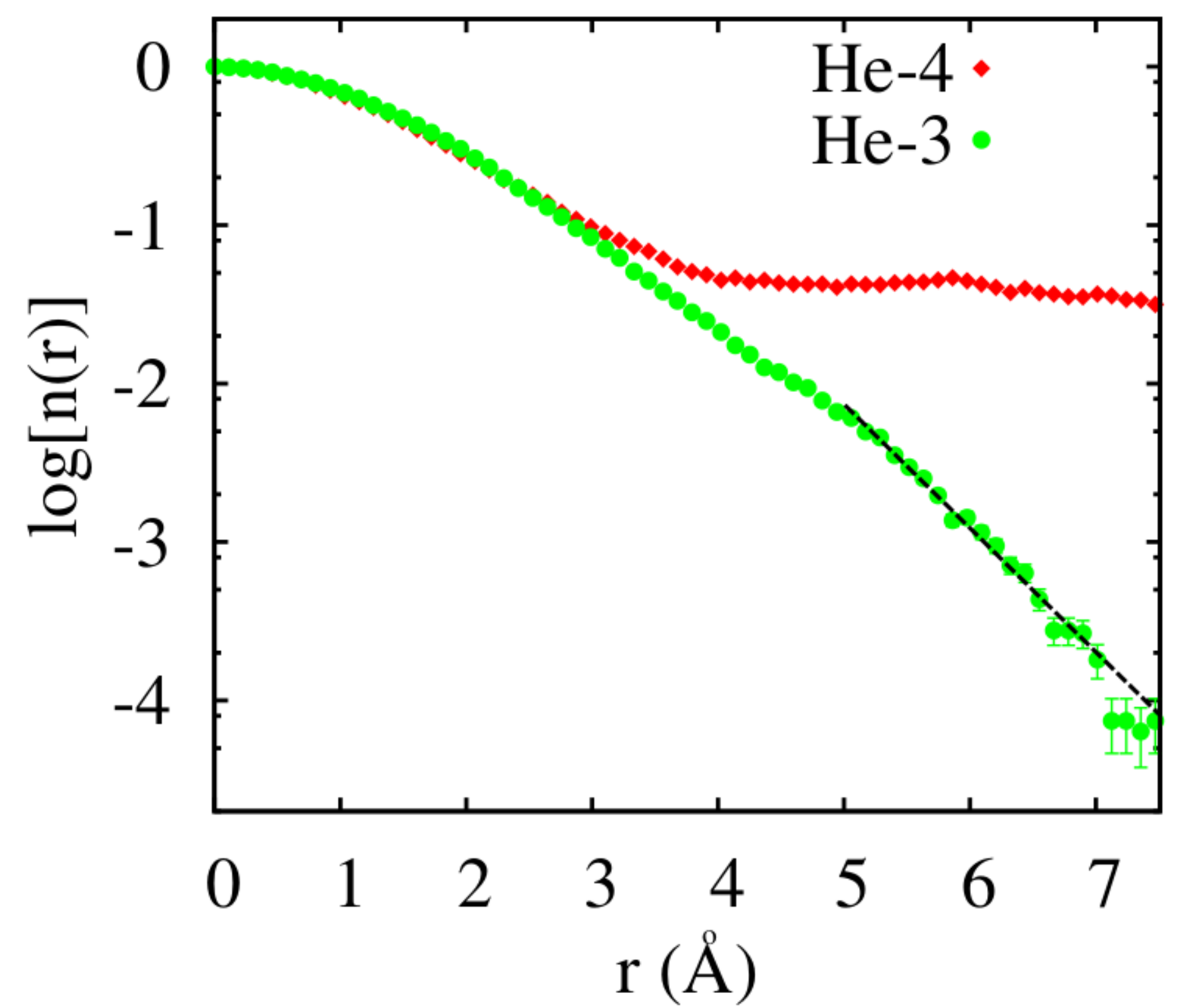}
\caption{{\rm Color online.} One-particle density matrix (log scale, base 10) computed by QMC for the $^4$He (diamonds) and $^3$He (circles) components of a mixture qith $x=10\%$, at $T$=2 K. The density of the mixture is 0.0214 \AA$^{-3}$. When not shown, statistical errors are smaller than the sizes of the symbols. Dashed line is an exponential fit to the $^3$He result for $r > 5$ \AA.}
\label{f2}
\end{figure}

Because the kinetic energy is experimentally determined through a measurement of the momentum distribution, we now turn to the discussion of this quantity in the mixture, specifically beginning with the  one-body density matrix. Fig. \ref{f2} shows the result for the same thermodynamic conditions of Fig. \ref{f1}. The two curves are nearly indistinguishable up to a distance of the order of the diameter of the repulsive core of the interatomic potential, as expected displaying markedly different behavior at long distances. 
\\ \indent
The $^4$He density matrix plateaus at long distance to a value slightly above 3\%, which is the estimate of the condensate fraction. 
This is approximately 25\% lower than the value in pure $^4$He at $T$=2 K, at the considerably higher equilibrium density $\rho_4=0.02194$ \AA$^{-3}$. The decrease of the condensate fraction is due to the presence of the $^3$He impurities, which have the effect of inhibiting in part long exchanges\cite{mb} of $^4$He atoms, which underlie both Bose condensation as well as the superfluid response. The $^4$He superfluid fraction $\rho_S$ is 0.17(3), which is consistent with the result quoted in Ref. \onlinecite {bm}, where the RPIMC was used. Its value in the pure $^4$He system at this temperature (at the aove mentioned density $\rho_4$) is close to 0.48. At these low $x$, the $^4$He density matrix and momentum distributions largely reproduce those for pure bulk $^4$He, extensively discussed elsewhere.\cite{rta} We therefore now focus on the $^3$He component.
\\ \indent 
The $n_3(r)$ shown in Fig. \ref{f2} has obviously a very different behavior from the $n_4$; the first obvious thing to notice is that, despite the neglect of quantum statistics, it 
is very different from a Gaussian, which is what it would be for a fluid of distinguishable quantum particles.\cite{mb09} 
It is monotonically decreasing, in a way that at large distances is consistent with an exponential decay, within the uncertainties of the calculation. Although a similar decay can be observed in the one-body density matrix of liquid parahydrogen at freezing, in that context it is due to quantum-mechanical exchanges; in this case, on the other hand, $^3$He atoms are regarded as truly distinguishable.  Rather, the exponential decay at long distances of the $n_3$ is consistent with the Landau-Pomeranchuk notion of a $^3$He atom penetrating a potential barrier, represented by the surrounding, nearly homogeneous superfluid $^4$He medium.
\\ \indent 
As noted above, the one-body density matrix often displays effects of quantum statistics that do not show up (as obviously) in structural or energetic properties of the system;\cite{mb09} thus, one need asses the possible effect of the neglect of quantum statistics on the results shown in Fig. \ref{f2}. One may expect deviations between the one-body density matrix computed by treating particles as distinguishable, and one in which Fermi statistics were taken into account, to appear at a distance of the order of the average separation between two exchange mates, i.e., two $^3$He atoms with parallel spin projections, which is close to 10 \AA\ at the physical conditions of the results of Fig. \ref{f2}; noting the exponentially decreasing behavior, and considering that the long-range part of the $n(r)$ affects the low-k part of the $f(k)$, we may conclude that whatever change Fermi statistics may impart to the $n_3(r)$ computed here, it is likely to have very little effect on the $^3$He kinetic energy.
\begin{figure}[h]
\centering
\includegraphics[width=\linewidth]{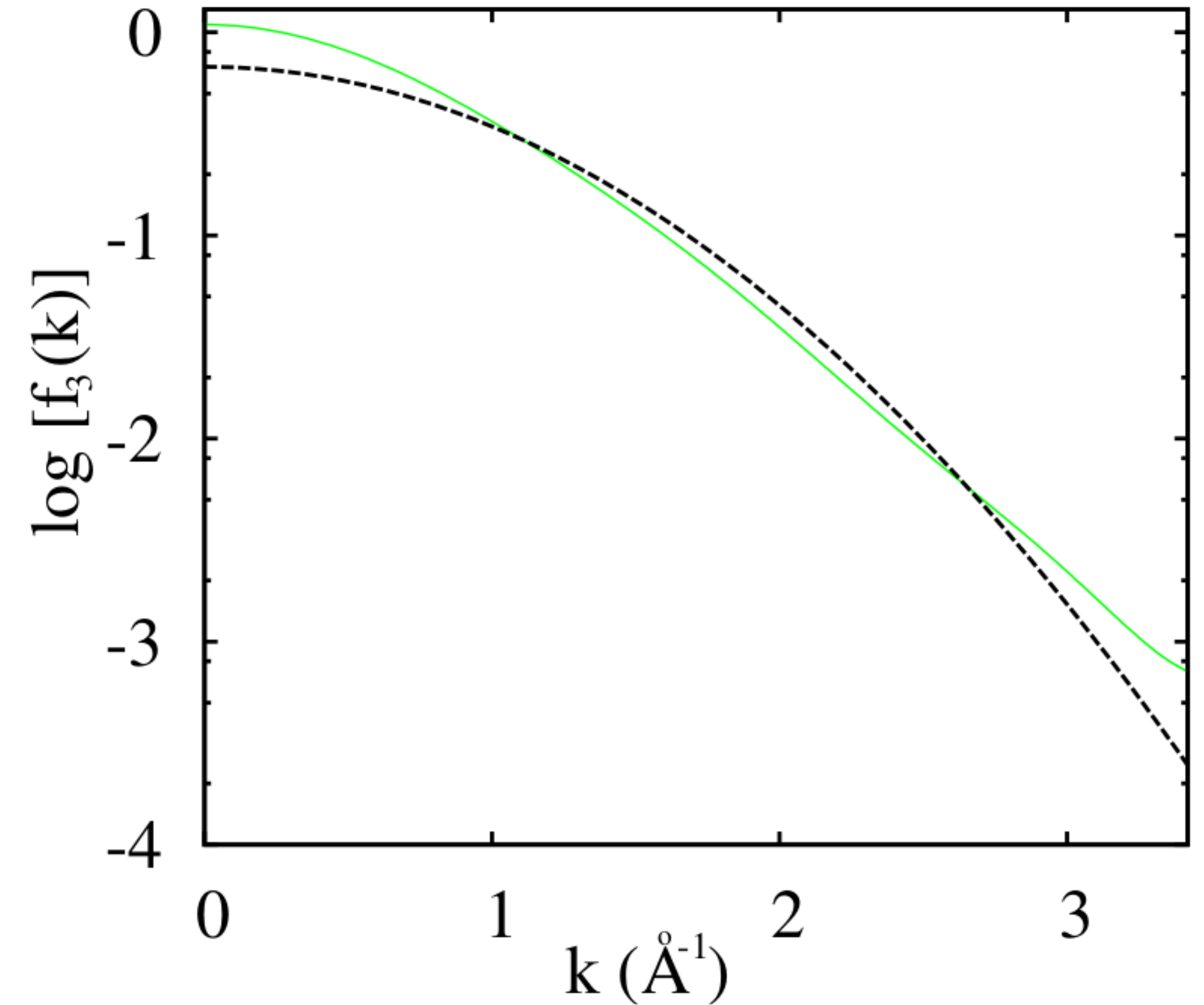}
\caption{{\rm Color online.} $^3$He momentum distribution $f_3(k)$ (solid line, log scale, base 10) for a liquid helium mixture with a $^3$He concentration $x=10\%$ at  temperature $T$=2 K (solid line). The function $f_3(k)$ is obtained using Eq. \ref{nofk} from the one-body density matrix $n_3(r)$ (shown in Fig. \ref{f2}) computed by QMC. The density of the mixture is 0.0214 \AA$^{-3}$. Statistical errors are not visible on the scale of the curve. Dashed line represents a Gaussian momentum distribution yielding the same value of the $^3$He kinetic energy per atom, namely 18.1(1) K.}
\label{f3}
\end{figure}

Fig. \ref{f3} shows the resulting. theoretically predicted momentum distribution $f_3(k)$ for the $^3$He component, obtained from the computed $n_3(r)$ through Eq. \ref{nofk}; specifically, a numerical integration was performed based on data up to a distance $r=7.5$ \AA, while the contribution from greater distances was evaluated analytically, based on an exponential fit of the data at long distance, as shown in Fig. \ref{f3}. Altogether, the contribution of the long-range part  of the $n_3(r)$ is not visible on the scale of the curve as shown in Fig. \ref{f3}. Also shown in Fig. \ref{f3}, for comparison, is a Gaussian model momentum distribution yielding the same kinetic energy per particle as the computed $f_3(k)$.
\\ \indent 
Obtaining accurate estimates for $f_3$ for momenta greater than $k \sim 3.4$ \AA$^{-1}$ using the above procedure, is rendered problematic by the discretization of $n_3(r)$. However, the result shown in Fig. \ref{f3} suffices to illustrate the main physical conclusions. The first obvious observation is significant deviation from a simple Gaussian, both at low momenta, where $f_3$ gains strength due to the enhanced delocalization of a $^3$He atom in superfluid $^4$He, as well as at high momenta, as a result of hard core repulsive interaction with nearby, heavy $^4$He atoms, which impart to the dissolved $^3$He atoms its renormalized mass. 
\\ \indent 
It is precisely the presence of this long, non-Gaussian tail in the $f_3(k)$ (note the logarithmic scale Fig. \ref{f3}), that renders the extraction of the kinetic energy from $f_3$ quite delicate, as already suggested by several authors.\cite{bc,diallo06} For, a substantial contribution to the kinetic energy comes from the tail; taking for example the two distributions shown in Fig. \ref{f3}, in the case of the Gaussian the portion of the distribution for momenta greater than 3 \AA$^{-1}$ contributes a mere 3\% of the total kinetic energy, but close to 20\% for the computed $f_3$. Thus, the use of an inadequate model to fit the experimentally measured distribution, especially one that does not properly describe the tail, can easily lead to an underestimation of $K_3$.
\section{conclusions}\label{conc}
We have carried out a computational study of isotopic liquid helium mixtures at a temperature $T$=2 K, with the aim of possibly shedding light on a present disagreement between theoretically computed and experimentally measured atomic kinetic energies. Our study is based on first principle computer simulations, whose only input is the interatomic pair potential; the only approximation built into our simulations is the neglect of quantum (Fermi) statistics for the $^3$He component. The results of the simulation confirm basic theoretical arguments suggesting that this is an excellent approximation for low $^3$He concentration (less than $\sim 20$\%), at the temperature considered here. However, the kinetic energy estimates obtained neglecting Fermi statistics seem reasonable even in the pure $^3$He limit. 
\\ \indent 
The momentum distribution for $^3$He at low $x$, where the disagreement between theory and experiment is most substantial, displays a slowly decaying tail at high momenta, arising from the short-range, repulsive interaction of a light $^3$He atom with the surrounding cage comprising heavy $^4$He atoms. This effect is expected to become progressively less important as the $^3$He concentration increases (and the equilibrium density correspondingly decreases). 
\\ \indent
It is worth noting that the non-condensate part of the momentum distribution of superfluid $^4$He, which is the one that contributes to the $^4$He kinetic energy, does not feature the same kind of long range tail,\cite{msf97}, but can actually fairly well be approximated by a Gaussian. This is why the determination of the kinetic energy is more accurate than for $^3$He in the low concentration mixture. On the other hand, as $x$ increases the $^4$He component turns normal, and concurrently the agreement between the computed and experimentally measured $K_4$ worsens (see result at $x=35\%$ in Table \ref{table1}), the experimental estimate again falling below the theoretical one.
\\ \indent 
In conclusions, this work supports the hypothesis first proposed in Ref. \onlinecite{bc}, subsequently expounded on in Ref. \onlinecite{diallo06}, that the disagreement between reported theoretical and experimental estimates of the kinetic energy of $^3$He in the mixture at low temperature, in the limit of low $^3$He concentration, may be the result of the model utilized to fit the measured momentum distribution.

\begin{acknowledgments}
This work was supported in part by the Natural Sciences and Engineering Research Council of Canada (NSERC). The author gratefully acknowledges the hospitality of the International Centre for Theoretical Physics, Trieste, where most of this research work was carried out. Computing support of Westgrid is also acknowledged.
\end{acknowledgments}

%\nocite{*}

\bibliography{mix}% Produces the bibliography via BibTeX.

\end{document}